\documentclass[superscriptaddress,amsmath,amssymb,aps,twocolumn,pra,notitlepage]{revtex4-2}
\usepackage{amsmath,amsfonts,amssymb,amsthm,graphics,graphicx,epsfig,bbm}
\usepackage[colorlinks=true,citecolor=blue,linkcolor=blue,urlcolor=blue]{hyperref}
\usepackage[usenames]{color}
\usepackage{graphicx}
\usepackage{subfigure}
\usepackage{amsmath}
\usepackage{epsfig}
\usepackage{dcolumn}
\usepackage{bm}
\usepackage{color}
\usepackage{times}
\usepackage{epstopdf}
\usepackage{amssymb}
\usepackage{amstext}
\usepackage{latexsym}
\usepackage{hyperref}
\usepackage{amsfonts}
\usepackage{psfrag}
\usepackage{soul,xcolor}
\usepackage[normalem]{ulem}
\usepackage{adjustbox}

\newcommand{\ket}[1]{\vert #1 \rangle}
\newcommand{\bra}[1]{\langle #1 \vert}
\newcommand{\ketbra}[2]{\vert #1 \rangle \langle #2 \vert}
\newcommand{\braket}[2]{\langle #1 \vert #2 \rangle}

\newcommand{\Tr}{\mathrm{Tr}}

\begin{document}

\title{Time evolution of nonlinear dynamics on a quantum processor}

\author{Jos\'e Diogo da Costa Jesus}
\affiliation{Forschungszentrum J\"ulich GmbH, Peter Gr\"unberg Institute, Quantum Control (PGI-8), 52425 J\"ulich, Germany}
\affiliation{Institute for Theoretical Physics, University of Cologne, D-50937 Cologne, Germany}

\author{Abhishek Setty}
\affiliation{Forschungszentrum J\"ulich GmbH, Peter Gr\"unberg Institute, Quantum Control (PGI-8), 52425 J\"ulich, Germany}
\affiliation{Institute for Theoretical Physics, University of Cologne, D-50937 Cologne, Germany}

\author{Tommaso Calarco}
\affiliation{Forschungszentrum J\"ulich GmbH, Peter Gr\"unberg Institute, Quantum Control (PGI-8), 52425 J\"ulich, Germany}
\affiliation{Institute for Theoretical Physics, University of Cologne, D-50937 Cologne, Germany}
\affiliation{Dipartimento di Fisica e Astronomia, Universit\`a di Bologna, 40127 Bologna, Italy}

\author{Dieter Jaksch}
\affiliation{University of Hamburg, Luruper Chaussee 149, 22761 Hamburg, Germany}
\affiliation{The Hamburg Centre for Ultrafast Imaging, Luruper Chaussee 149, Hamburg D-22761, Germany}
\affiliation{Clarendon Laboratory, University of Oxford, Parks Road, Oxford OX1 3PU, United Kingdom}

\author{F. A. C\'ardenas-L\'opez}
\affiliation{Forschungszentrum J\"ulich GmbH, Peter Gr\"unberg Institute, Quantum Control (PGI-8), 52425 J\"ulich, Germany}

\author{Felix Motzoi}
\affiliation{Forschungszentrum J\"ulich GmbH, Peter Gr\"unberg Institute, Quantum Control (PGI-8), 52425 J\"ulich, Germany}
\affiliation{Institute for Theoretical Physics, University of Cologne, D-50937 Cologne, Germany}

\date{\today}

\begin{abstract}
    From fluid flow and transport to collective dynamics, numerical simulation of nonlinear partial differential equations underpins modern scientific computing. Extending this capability to quantum computers remains a longstanding challenge because nonlinear and non-Hermitian evolution is fundamentally incompatible with conventional Hamiltonian-based quantum simulation. Here we experimentally realize the time evolution of nonlinear fluid dynamics on a quantum processor using a hybrid variational framework for the viscous and inviscid Burgers equations. Our approach directly encodes the nonlinear dynamics into a variational optimization procedure, avoiding the enlarged linear embeddings and truncation overhead associated with Carleman linearization-based quantum algorithms. We further demonstrate convection-dominated dynamics corresponding to Reynolds numbers of order $10^2$. We encode the governing evolution into parametrized quantum circuits and iteratively reconstruct the time-dependent field through quantum–classical optimization. By introducing a zero-noise extrapolation method without additional circuit-folding overhead, we accurately execute deep error-mitigated circuits with entangling-gate counts beyond those typical of Hadamard test circuits. We accurately reconstruct the time evolution across multiple timesteps despite hardware noise and finite device coherence. Our results constitute, to our knowledge, the first experimental realization of nonlinear time propagation on a quantum processor, extending quantum simulation beyond predominantly linear settings and establishing a route toward quantum computation for nonlinear continuum dynamics. 
    
\end{abstract}

\maketitle

\section*{Introduction}

Nonlinear partial differential equations (PDEs) govern a broad range of phenomena across physics, engineering, and the natural sciences, including but not limited to fluid flow, transport, turbulence, reaction--diffusion dynamics, and collective emergent behaviour~\cite{ames1965nonlinear,myint2007linear,logan2008introduction}. Understanding the real-time evolution of such systems is central to fields ranging from computational fluid dynamics (CFD) and plasma physics to climate modelling and nonequilibrium statistical mechanics~\cite{anderson1995computational,wendt2008computational}. Quantum computation has emerged as a promising paradigm for simulating complex dynamical systems due to its ability to efficiently encode exponentially large state spaces in quantum superposition states, allowing for the vast grid sizes required to simulate such complex phenomena ~\cite{berry2014high,tosti2022review}. However, most existing quantum simulation frameworks are intrinsically tailored to linear and unitary dynamics, whereas many physically relevant continuum systems are fundamentally nonlinear and effectively non-Hermitian. As a result, the realization of nonlinear continuum dynamics on quantum hardware remains a longstanding challenge.

Substantial theoretical progress has been made toward quantum algorithms for differential equations, including quantum linear system solvers~\cite{harrow2009quantum,berry2014high,setty2025block,setty2026quantum,liu2021efficient}, variational and Hadamard-test-based formulations~\cite{lubasch2020variational,jaksch2023variational,over2025boundary}, physics-informed quantum machine learning approaches~\cite{setty2025self,kyriienko2021solving}, and spectral methods~\cite{childs2020quantum}. Among these, variational formulations based on spatial discretization are particularly appealing for continuum dynamics because they naturally connect to established finite-difference and finite-volume approaches used in classical numerical simulation. These are especially suited for current devices when compared with other state approaches, which require much larger circuits~\cite{setty2025block,mcclean2016theory,Cerezo2021, bengoechea2026quantum} . In particular, the framework introduced by Lubasch \emph{et al.}~\cite{lubasch2020variational} provides a systematic route to encode discretized differential operators into parametrized quantum circuits by mapping field amplitudes onto the amplitudes of a quantum state. This enables exponential state-space encoding and efficient evaluation of inner products between quantum states. Importantly, nonlinear terms can be incorporated through extended circuit constructions, enabling simulation strategies that go beyond the intrinsic linearity of quantum mechanics while avoiding the enlarged Hilbert-space embeddings and truncation errors characteristic of Carleman linearization approaches \cite{setty2025block,PhysRevResearch.7.023254, Krovi_2023}. Despite these advances, experimentally realizing nonlinear time evolution remains difficult. Non-Hermitian generators typically require additional ancillary resources, nonlocal operations, intermediate measurements, or enlarged Hilbert-space embeddings~\cite{Fleckenstein2022}, leading to substantial circuit overhead and increased optimization complexity. Furthermore, variational implementations often require additional trainable parameters to relax normalization constraints or stabilize evolution~\cite{lubasch2020variational}, producing increasingly intricate optimization landscapes that are difficult to navigate in the presence of hardware noise.

These challenges are particularly acute on present-day noisy  processors~\cite{Bharti2022}, where finite coherence times, gate imperfections, and limited measurement budgets constrain the accessible circuit depth and algorithmic fidelity. Although current devices provide sufficient qubit counts to encode discretized fields, circuit depth remains the principal bottleneck limiting experimental realizations of nonlinear dynamical simulation. In particular, iterative time propagation compounds both coherent and stochastic errors, while the interplay between trainability, barren plateaus, and error mitigation in variational dynamics remains poorly understood~\cite{wang2024can,botelho2022error,quek2024exponentially,robbiati2026real,cai2023quantum}. Recent experimental efforts have begun exploring quantum fluid dynamics in linear or Hamiltonian-based settings. For example, Chen \emph{et al.}~\cite{chen2024enabling} implemented Poiseuille flow using a variational quantum linear solver~\cite{bravo2023variational}, while Meng \emph{et al.}~\cite{meng2024simulating} simulated vortex dynamics through Hamiltonian evolution on superconducting quantum hardware. However, experimental realizations of nonlinear time-dependent continuum dynamics remains limited due to very deep variational circuits, also ruling out most error mitigation approaches.
\begin{figure*}[!t]
    \centering
    \includegraphics[width=0.9\textwidth]{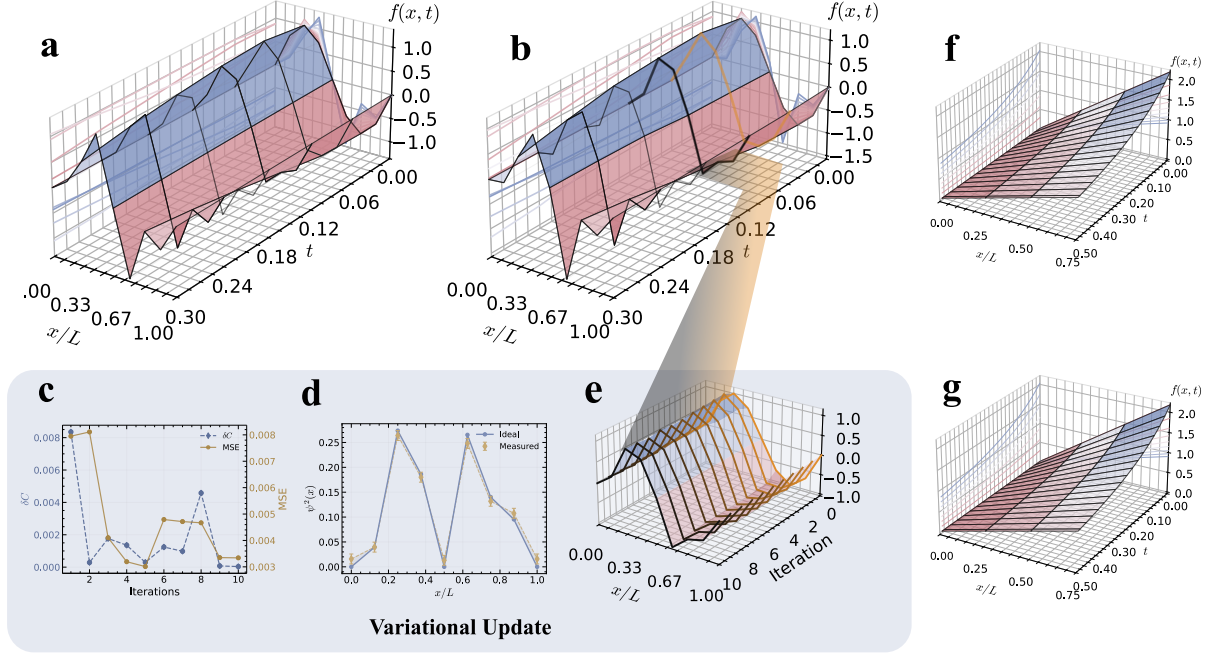}
    \caption{
\textbf{Evolution of the Burgers equation in a convection-dominated regime.}
\textbf{a)} The ideal evolution of the viscous Burgers equation with viscosity $\nu=0.01$ for the initial condition $f(x,0)=\sin(2\pi x)$. The competition between nonlinear convection and viscous diffusion progressively generates sharp spatial gradients and shock-like structures during the evolution. The evolution is performed variationally with timestep $dt=0.06$ up to final time $T=0.3$. 
\textbf{b)} Experimentally reconstructed evolution obtained using the variational quantum time-propagation protocol (see Figure \ref{fig:dynamics}), showing close agreement with the ideal nonlinear dynamics. The final MSE is 2.1\% compared to the error-free dynamics, while the reconstructed intermediate states 
for each time step have MSE of 0.3\%, 0.4\%, 0.9\%, 1.5\% and 2.1\% and state-overlap fidelity of 99.9\%, 99.3\%, 98.3\%, 97.3\% and 96.1\%, respectively.
\textbf{c)}  Optimization convergence for the second timestep, showing the evolution of the cost-function difference $\delta C$ and the MSE between the optimized quantum state and the ideal reference solution during the SGEO optimization procedure.
\textbf{d)} Measurement of the square of the field at the end of the variational update.  It is obtained directly (i.e.~without quantum state tomography) by reading out the ansatz circuit statistics, showing excellent agreement with the numerically computed ideal states. 
\textbf{e)} Evolution of the field for a single time step over the SGEO optimization procedure, demonstrating the warm start and a quick convergence towards the desired state.  
\textbf{f)} Ideal evolution of the inviscid Burgers equation for the initial condition $f(x,0)=3x$. In the absence of viscosity, the dynamics are governed entirely by nonlinear convection, leading to progressive wave steepening without diffusive regularization.
\textbf{g)} Inviscid evolution obtained from the hardware simulation to compute the norm at each time step, reusing optimal parameters. These reproduce the transport-driven nonlinear dynamics with MSE of 0.01\%, in close agreement with the ideal solution.
    }
    \label{fig:timevol}
\end{figure*}

Here, we experimentally realize nonlinear fluid dynamics on superconducting quantum processors using a hybrid quantum--classical variational framework for the viscous and inviscid Burgers equations in a convection-dominated regime corresponding to Reynolds numbers of order $10^
2$, where nonlinear steepening strongly competes with diffusion. We implement iterative time propagation using parametrized quantum circuits encoding the governing differential operators and benchmark multiple optimization strategies, including COBYLA and Sequential Grid-based Explicit Optimization (SGEO)~\cite{umer2025efficient, mastorakis2026resource}, finding enhanced robustness of SGEO in noisy settings. To enable reliable execution of deep variational circuits, we combine gate-decomposition-aware ansatz design~\cite{barenco1995background} with error mitigation techniques including zero-noise extrapolation~\cite{giurgica2020digital} and a novel approach to inverting depolarizing error in Hadamard circuits while avoiding circuit folding.
This enables our implementations to reach circuit depths containing up to 60 controlled-$Z$ layers on IBM superconducting quantum hardware with heavy-hex connectivity while retaining nearly unit algorithmic fidelity. We further investigate warm-start initialization protocols and introduce a method for estimating solution norms directly from quantum measurements without requiring full state tomography. Together, our results demonstrate nonlinear time evolution on quantum processors across multiple timesteps despite finite device coherence and hardware noise, extending quantum simulation beyond predominantly linear and unitary regimes and establishing a route toward quantum computation for nonlinear continuum dynamics.

\section*{Results}

\begin{figure*}[!t]
    \centering
    \includegraphics[width=1.0\linewidth]{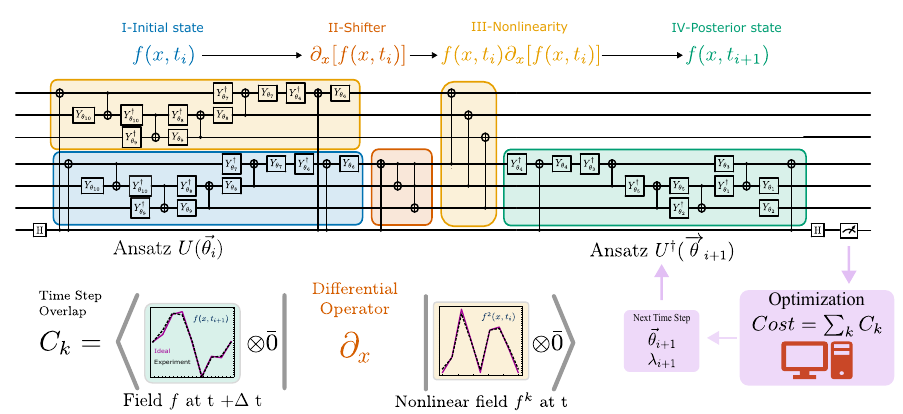}
    \caption{
\textbf{Variational circuit decomposition for nonlinear Burgers dynamics.}
 Quantum circuit implementing the convection term  of the Burgers equation. Highlighted circuit components correspond to distinct contributions entering the nonlinear evolution. The blue block encodes the parametrized quantum state representing the field $f(x,t)$ at the current timestep. The red block implements the discretized shifting operator associated with the spatial derivatives (a finite central derivative requires two shifters-- one forwards and one backwards, see Eq. \ref{eq:dx_operator}). The orange blocks represents the nonlinear-processing-unit (QNPU) which evaluates products of field amplitudes required for the nonlinear convective term $f\,\partial_x f$. The teal block encodes the posterior variational state representing the updated field at time $t+\Delta t$, whose parameters are optimized through the variational propagation procedure. Measurements from this circuit are used to construct the Time Step Overlap $C_k$ between the evolution of the previous state and the posterior state. The sum of all $C_k$ generates the cost function which is minimized using classical computing. This provides the parameters $\lambda_{i+1}$ and $\overrightarrow{\theta}_{i+1}$ for the next time step. }
    \label{fig:dynamics}
\end{figure*}
\subsection*{Variational encoding of nonlinear dynamics}

Conventional quantum simulation naturally implements linear unitary evolution, whereas nonlinear partial differential equations require state-dependent dynamics that are generally non-unitary, not norm-preserving and, after discretization, effectively non-Hermitian. Existing quantum approaches often address this mismatch through enlarged linear embeddings, such as Carleman linearization, which reformulate nonlinear evolution as a higher-dimensional linear problem at the cost of truncation errors and substantial auxiliary-resource overhead. Here, instead, we formulate the dynamics directly as a variational consistency problem between successive quantum states, where nonlinear contributions are represented intrinsically through amplitude-level products and higher-order state overlaps, rather than through a linearized embedding of the dynamics.

As a paradigmatic example, we consider the Burgers equation,
\begin{equation}
\frac{\partial}{\partial t} f(x,t)
= \nu \frac{\partial^2}{\partial x^2} f(x,t)
- f(x,t)\frac{\partial}{\partial x} f(x,t),
\end{equation}
which describes the competition between viscous diffusion, which smooths spatial structure, and nonlinear convection, which dynamically steepens the field profile and drives shock formation. The spatial domain of the field is discretised using $n$ qubits corresponding to $2^n$ grid points. Thus, the discretized field $f(x,t)$ is encoded into the amplitudes of a parametrized quantum state,
\begin{equation}
\lambda|\psi(\overrightarrow{\theta}(t))\rangle
= \lambda \hat{U}_{\mathrm{enc}}(\overrightarrow{\theta}(t))|\bar{0}\rangle,
\end{equation}
where the variational parameters $\overrightarrow{\theta}(t)$ determine the encoded field amplitudes and $\lambda$ captures the overall normalization. The field encoding circuit consists in layered single-qubit rotations and entangling gates with a total depth of $d$ and approximately $N$ elementary gates per timestep optimization. We should note that the Burgers dynamics are not norm preserving; this normalization must be updated at every timestep alongside the quantum state parameters. The constant monitoring of this parameter is fundamental to achieve successful time propagation of a non-normalized partial differential equation on a quantum processor.

Time evolution is formulated not as unitary propagation but rather as a sequence of variational consistency steps between consecutive timesteps. At each step, $\overrightarrow{\theta}(t+\Delta t)$ is obtained by minimizing a cost function constructed from overlaps between powers of parametrized quantum states and operator-encoded spatial derivatives. Within this framework, nonlinear contributions, in particular the convective term $f\,\partial_x f$, arise naturally from products of amplitudes encoded in higher-order overlap constraints evaluated via Hadamard-test interferometry. Importantly, this avoids any explicit linearization of the nonlinear field dynamics. Further details regarding how the different operators are encoded into circuits can be found in Methods \ref{Algo_Comp_Methods}.

\subsection*{Experimental realization of nonlinear Burgers dynamics}

We experimentally implement the variational time-propagation framework for both the viscous and inviscid Burgers equations on superconducting quantum hardware. 

Figure~\ref{fig:timevol} summarizes the experimentally reconstructed dynamics. In the ideal noiseless evolution (Fig.~\ref{fig:timevol}a), the initially smooth sinusoidal profile is evolved in time. For $\nu = 0.01$, corresponding to an effective Reynolds number $Re \approx 100$ for the chosen initial condition, the dynamics lie in a convection-dominated regime in which nonlinear steepening competes directly with diffusion, developing strongly localized sharp spatial gradients.

The corresponding hardware evolution (Fig.~\ref{fig:timevol}b) closely tracks the ideal dynamics across all timesteps, reproducing both waveform steepening and the emergence of localized gradient structures. The final reconstructed field reaches a mean-squared error (MSE) of approximately $2\%$ relative to the exact solution, demonstrating stable iterative propagation despite hardware noise. This indicates that the variational feedback loop maintains dynamical consistency over multiple timesteps beyond single-circuit operation.

The update of the field for each time step follows the optimization dynamics shown in the blue box of Figure~\ref{fig:timevol}. The SGEO optimizer exhibits rapid convergence to the new state within ten iterations when initialized with a warm start from the previous time step's solution. 

This reflects a quite smooth variational landscape between consecutive timesteps for sufficiently small $\Delta t$, enabling efficient iterative propagation without extensive optimization overhead. Given the proximity from the initital to optimized state, such a warm start can be found by picking a small enough $\Delta t$ \cite{2qzh-yf49}. 
During optimization, the reconstructed field (Fig.~\ref{fig:timevol}d) evolves from the previous timestep toward the target solution, with progressive sharpening of gradients consistent with nonlinear convective transport and viscous regularization. This evolution directly reflects the emergence of the physical solution within the variational parameter space.

\begin{figure*}[t!]
    \centering
    \includegraphics[width=1\linewidth]{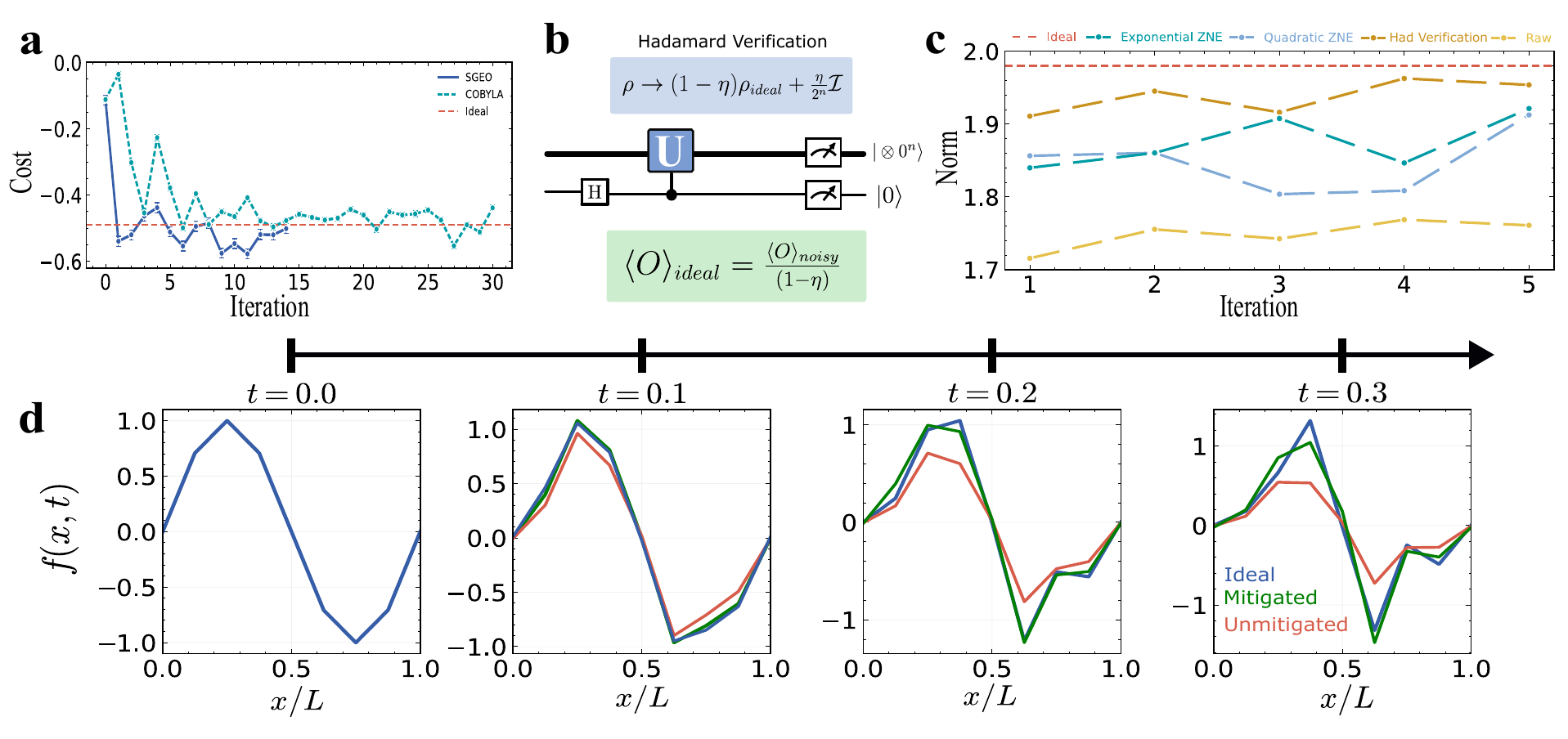}
    \caption{ \textbf{ Optimization and Error Mitigation Strategies for the Burgers dynamics considered} \textbf{a)} Comparison between SGEO and COBYLA optimization for the Inviscid problem, for a single time step dt=0.1 and a cold start. Although COBYLA finds a good minima for the cost, it cannot sufficiently resolve the $\overrightarrow{\theta}$ parameters and produces incorrect solutions. For this single time step the solution obtained via COBYLA has a MSE of 0.026 and a fidelity of 94.2\% and the solution obtained via SGEO has a MSE of 0.01 and a fidelity of 99.2 \% . \textbf{b)} Schematic representation of the Hadamard Verification method used for error mitigation. The controlled circuit $U$ is assumed to produce a global depolorizing noise channel. By not applying the last Hadamard qubit and measuring all qubits, the global depolorizing noise factor $\eta$ can be estimated by comparing the number of times all qubit are measured in $\ket{0}$ and the number of times the Hadamard qubit is measured at $\ket{0}$. \textbf{c)} Comparison between different error mitigation methods for norm extraction. For a single time step, 5 measurements of the norm were performed. Hadamard Verification outperforms all error mitigation methods (Quadratic and Exponential ZNE), with all methods outperforming the raw result without any mitigation. This indicates the deepest circuits of Zero Noise Extrapolation become to long and produce completely scrambled results. \textbf{d)} Time-marching for the Viscid Burgers Equation for 3 time steps, each with dt=0.1, using Error Mitigation for the norm extraction. The results obtained show fidelities of 99.8\%, 99.0\% and 95.7\%  and MSE of 0.009, 0.001 and 0.024 respectively (compared with MSE of 0.013, 0.056 and 0.129 for the unmitigated case, respectively).} 
    \label{fig:Time_March_Mit}
\end{figure*}

To further assess generality, we consider the inviscid Burgers equation, where the diffusive term is absent and the dynamics are purely convective. Fig.~\ref{fig:timevol}f and g show the ideal and hardware-reconstructed evolution for an initially linear profile, respectively. The solution undergoes slope deformation induced by nonlinear transport. The hardware results remain in agreement with the ideal evolution throughout the propagation, achieving a MSE of 0.01\%. Fig.~\ref{fig:Time_March_Mit}a shows the cost evolution for the invisicid Burgers equation, starting from a cold start. SGEO sucessfully minimizes the cost, finding parameters with fidelity of $99.2$ \% when compared to the ideal solution. 

Overall, these results demonstrate that present-day superconducting quantum processors can faithfully encode and propagate nonlinear continuum dynamics within a unified variational framework, capturing both viscous shock formation and inviscid nonlinear transport.

\subsection*{Variational circuit construction and dynamical encoding}

The nonlinear timestep update underlying the Burgers evolution is decomposed into experimentally measurable circuit primitives. Different components of the cost function will be associated with different operators of the equation. Figure~\ref{fig:dynamics} illustrates the convection term: the term including the non-linearity and the deepest circuit. The quantum circuit separates the propagation into distinct components associated with the encoded field, spatial derivative operators, nonlinear field products, and the posterior variational state after propagation. Two copies of the parametrized quantum state encode the discretized field $f(x,t)$, while additional operator blocks implement the viscous Laplacian and nonlinear convective contributions entering the Burgers equation. Methods\ref{Algo_Comp_Methods} goes into further details on how spatial derivatives are transformed into quantum circuits and all required quantum operators to solve the viscous and inviscid Burgers equation.

The quantum nonlinear-processing-unit (QNPU) block evaluates amplitude-level products between encoded quantum states required for the nonlinear convection term $f\,\partial_x f$. In contrast to approaches based on Carleman linearization, where nonlinear dynamics are embedded into enlarged linear systems with truncated auxiliary hierarchies, the present approach generates nonlinear field products directly through higher-order quantum overlaps between parametrized states. Nonlinear evolution is therefore implemented intrinsically at the wavefunction level, the no-cloning theorem is circumvented by enforcing a gridpoint-selective product using CNOT gates conditioned on the first Ansatz state (see beige circuit in Figure ~\ref{fig:dynamics}). 

The variational timestep update is obtained by minimizing a cost function of the form
\begin{equation}
\mathcal{C}(\lambda_{i+1},\overrightarrow{\theta}_{i+1})
=
|\lambda_{i+1}|^2
+
|\lambda_{i}|^2
-
2\lambda_{i+1}\lambda_{j}
\sum_k
\mathcal{C}_k(\overrightarrow{\theta}_{i+1}),
\label{eq:main_cost}
\end{equation}
where the overlap contributions $\mathcal{C}_k$ encode the discretized diffusion and convection operators entering the Burgers dynamics and $i$ indicates the time step.
Minimization of the cost function determines the posterior variational state representing the propagated field at timestep $t+\Delta t$.

Each overlap contribution is evaluated through Hadamard-test interferometry of the form
\begin{equation}
\langle \psi(\overrightarrow{\theta}_{t+\Delta t}) |
\hat{\mathcal O}_k
|
\psi(\overrightarrow{\theta}_{t})
\rangle,
\end{equation}
where $\hat{\mathcal O}_k$ denotes derivative and nonlinear operator components of the discretized evolution. Figure~\ref{fig:dynamics} further shows experimentally reconstructed encoded fields, nonlinear field products generated within the QNPU block, and the optimized posterior variational state after propagation. The close agreement between experimental measurements and the ideal results demonstrates accurate implementation of both the diffusive and nonlinear convective components of the Burgers dynamics on superconducting quantum hardware.

Because the nonlinear operator blocks require deeper controlled-overlap evaluations, stable propagation further relies on error-mitigated circuit execution, including dynamical decoupling and mitigation applied directly to the overlap measurements entering the variational update.

\subsection*{Error mitigation and stability of iterative nonlinear propagation}

In contrast to conventional unitary quantum simulation, where hardware noise primarily perturbs measured observables, errors in variational nonlinear dynamics can feed back directly into the time-propagation procedure itself. Because each timestep is reconstructed from variational overlaps with the previous state, small inaccuracies can accumulate nonlinearly across successive updates, biasing both the reconstructed field and its normalization. This effect is particularly important because the normalization parameter $\lambda_i$ enters explicitly into the variational cost function and therefore directly influences the stability of subsequent timesteps. Error mitigation therefore becomes an intrinsic component of the nonlinear propagation algorithm rather than a post-processing correction applied after measurement.

To stabilize the iterative evolution, we compare zero-noise extrapolation with a novel depolarization inversion postselection approach 
capable of mitigating errors in the overlap measurements entering the Hadamard-test cost function. Unlike approaches requiring circuit folding or additional symmetry-enforcement layers, the noise-inversion protocol operates directly at the level of expectation-value estimation without substantially increasing circuit depth. The output of the Hadamard test is thus rescaled with an estimate of the circuit depolarization, which we label $\eta$. The depolarization is estimated via the residual population outside the ground state when the Hadamard qubit is itself in the ground. This estimate then provides a streamlined way to invert the depolarization without code folding, giving 
\begin{equation}
    \langle O \rangle_{ideal} = \frac{\langle O \rangle_{noisy}}{(1-\eta)}
\end{equation}

In Fig.~\ref{fig:Time_March_Mit}b we show a schematic of this protocol, that we term Hadamard Verification, and in Fig.~\ref{fig:Time_March_Mit}c we compare the different error mitigation protocols when running a single norm extraction for the viscous Burgers equation, with a time step of $dt=0.1$.

This becomes crucially important for the nonlinear operator blocks, whose controlled overlap measurements require significantly deeper entangling circuits than the linear contributions and are therefore more susceptible to decoherence and accumulated gate error, as well as information scrambling~\cite{PhysRevX.9.011006,PhysRevX.11.021010,Landsman2019}.

The impact of mitigation on iterative nonlinear propagation is shown in Fig.~\ref{fig:Time_March_Mit}d. Without mitigation, the reconstructed evolution initially follows the expected Burgers dynamics but progressively develops a compounded systematic drift in the normalization parameter $\lambda_j$. Because the variational update at each timestep depends recursively on the reconstructed state from the previous step, this normalization error feeds back into the subsequent evolution and eventually drives an artificial collapse of the encoded field amplitudes. In contrast, the mitigated evolution remains stable throughout the full propagation, preserving both the nonlinear deformation of the field and the correct norm evolution over multiple timesteps.

Quantitatively, the mitigated evolution achieves a final-state  fidelity of approximately $95.7\%$ with a MSE of $0.024$, compared with $93.0\%$ and $0.129$ respectively in the unmitigated case. More significantly, the reconstructed normalization remains within $1\%$ of the ideal value under mitigation, whereas the unmitigated evolution accumulates deviations exceeding $80\%$. These results demonstrate that the dominant failure mechanism in iterative nonlinear quantum simulation is not simply local gate noise, but the recursive amplification of normalization errors through the variational feedback loop. Stabilizing this feedback is therefore essential for achieving physically meaningful long-time nonlinear dynamics on near-term quantum hardware.

\section*{Discussion}

We have demonstrated nonlinear time propagation on a superconducting quantum processor, establishing an experimental route toward quantum simulation of nonlinear continuum dynamics. In contrast to approaches based on enlarged linear embeddings, such as Carleman linearization, the present method implements nonlinear dynamics directly through variational overlap constraints between successive quantum states. Nonlinear field interactions are therefore encoded intrinsically through higher-order amplitude correlations and quantum interference measurements, avoiding the truncation hierarchy and auxiliary-state overhead associated with linearized formulations.

Using the viscous and inviscid Burgers equations as paradigmatic examples, we experimentally realized iterative nonlinear time propagation on near-term quantum processors in both diffusive and convection-dominated regimes. In the viscous case, the simulations accessed dynamics corresponding to an effective Reynolds number of approximately $Re\approx100$, where nonlinear convective steepening competes strongly with diffusion and generates shock-like structures during the evolution. The experimental results demonstrate that variational quantum propagation remains stable even in regimes where steep spatial gradients emerge dynamically throughout the time evolution.

A central observation of this work is that nonlinear quantum simulation introduces a qualitatively distinct error mechanism compared with conventional unitary quantum dynamics. Because each timestep is reconstructed recursively from variational overlaps with previous states, noise propagates through the nonlinear feedback loop and can destabilize the encoded norm governing the time-marching procedure. Error mitigation therefore plays a functional role in stabilizing the nonlinear dynamics itself rather than merely improving instantaneous circuit fidelity. By combining variational propagation with mitigation-aware overlap estimation, we achieved stable multi-step evolution with a final MSE of 0.021 for 5 time steps and norm errors of $\approx1\%$.

The present implementation constitutes a proof of principle for hybrid quantum--classical simulation of nonlinear dynamics on near-term hardware. Although the demonstrations presented here involve few-qubit discretizations, the framework itself is general and naturally extensible to larger spatial grids, higher-dimensional systems, and more complex nonlinear equations. Moreover, with a small number of ancilla qubits, the method is able to treat significant nonlinearity beyond what can be achieved with linearization methods with similar number of qubits. The variational formulation can incorporate arbitrary polynomial nonlinearities through generalized overlap constructions, making the approach applicable beyond fluid dynamics to nonlinear wave equations, reaction--diffusion systems, and nonlinear transport phenomena more broadly.

Looking forward, the next generation of quantum machines promises to step into the realm of early fault tolerance \cite{IBMRoadmap, QuantinuumRoadmap, IQMRoadmap, QuEraRoadmap}. In this regime, although qubit counts are still limited and 2 qubit gate errors are expected to be in the order of $\approx 0.01$\%, partial error correction becomes available. For instance, Clifford gates can be then implemented in a fault tolerant way while the single qubit rotations can be implemented via techniques like magic state injection \cite{lao2022magic}, permitting significant improvements in circuit fidelities and therefore longer circuits \cite{dangwal2025variational}.
When compounded with error mitigation strategies, one can expect even further improvements and larger experiments. 

Several important directions remain for future work. On the algorithmic side, improved ansatz constructions and adaptive timestep strategies may further enhance the stability and scalability of iterative propagation. Improvements in variance reduction \cite{preti2025gradients}, in encoding more versatile geometries, and in tailored variational optimization methods will further benefit the algorithm. More broadly, the present results provide a natural experimental framework for implementing nonlinear dynamical evolution directly at the level of quantum-state interference and amplitude correlations, and demonstrating a new route towards quantum simulation of nonlinear continuum physics.

\section*{Acknowledgements}

We thank Manuel Guatto, Dimitrios Georgiadis, Sahil Ugale, and Nikkin Deveraju for valuable discussions and insightful feedback. This work was supported by the European Union’s Horizon Europe programme through the QCFD project (Grant No. 101080085, HORIZON-CL4-2021-DIGITAL-EMERGING-02-10),  OpenSuperQPlus100 (Grant No. 101113946, HORIZON-CL4-2022-QUANTUM-01-SGA), and PASQuanS2.1 (Grant No. 101113690, HORIZON-CL4-2022-QUANTUM-02-SGA), the ML4Q2 Cluster of Excellence (EXC 2004/2 – 390534769), the German Federal Ministry of Education and Research (BMBF) through the QSolid project (Grant No. 13N16149), and the Helmholtz Initiative and Networking Fund projects KA-QUS-02 (qFLOW) and KA-QUS-03 (QT-Batt). The authors also acknowledge QuTech and IBM Quantum services for access to quantum computing resources. This research utilized resources of the Oak Ridge Leadership Computing Facility at Oak Ridge National Laboratory, supported by the U.S. Department of Energy, Office of Science, under Contract No. DE-AC05-00OR22725. All data used to generate the figures presented in this work are publicly available in Ref.~\cite{Repo}.

\section*{Methods}

\subsection{Algorithm and Equation Components}\label{Algo_Comp_Methods}\setcounter{subsection}{1}
We follow the approach developed in Ref.~\cite{lubasch2020variational, jaksch2023variational}, where the different terms of the PDE are implemented in terms of variational circuits consisting on a unitary matrix $\hat{U}(\overrightarrow{\theta})$ parametrized in terms of rotation angles $\overrightarrow{\theta}$, and these angles are obtained through a hybrid classical-quantum algorithm~\cite{cerezo2021variational} that optimizes a cost function $\mathcal{C}(\overrightarrow{\theta})$. The field is obtained by measuring an auxiliary system via a Hadamard test. Therefore, to encode a PDE on a quantum computer we requires two ingredients: the ansatz implementing $\hat{U}(\overrightarrow{\theta})$ and the Quantum Non-Linear Processing Unit (QNPU) that implements the different terms of the differential equation. 

We show how to encode the different terms of a PDE on variational circuits, and analyse how the operations required to do so scales with the number of two-qubit gates, which in current quantum processors are the limiting factor at the moment to execute a quantum circuit. Let us consider a general form of a non-linear partial differential equation:
\begin{equation}
    \frac{\partial}{\partial t}f(x,t) =  \mathbf{\hat{O}}(f(x,t)) f(x,t),
    \label{eq:general_pde}
\end{equation}
where $\mathbf{\hat{O}}$ represents any possible combination of power of the field and its derivatives:
\begin{equation}
    \mathbf{\hat{O}}(f(x,t)) = \sum_{i,j,k} c_{i,j,k}f^{i}(x,t)\frac{\partial^j }{\partial x^j}f^{k}(x,t).
    \label{eq:oper_O}
\end{equation}
The discretized field $f(x,t)$ is amplitude encoded into an n-qubit variational state. Here, we follow the amplitude encoding approach, where every component of the wavefunction $\ket{\psi^i_{\ell}}$ will corresponds to a position component $x_{\ell}$ of the field at time $t_i$. Thus, for a system of $N$ qubits, the spatial discretization will corresponds to $d=2^{N}$. To encode the time in this problem, we update the $\overrightarrow{\theta}$ at different time steps, such that each time $t_i$ is associated with its set of parameters $\overrightarrow{\theta}_i$. 

Moreover, different to quantum systems, the field $f(x,t)$ is not normalized. Thus, we are required to add an additional parameter $\lambda_i$. The discretization rules is given as 
\begin{equation}
    f(x,t) \equiv f(x, t_i) = \lambda_i \hat{U}_{{\rm{enc}}}(\overrightarrow{\theta}_i)|\mathbf{\bar{0}}\rangle,
    \label{Encoding}
\end{equation}
where $\ket{\mathbf{\bar{0}}}=\bigotimes_{\ell}\ket{0_{\ell}}$ is the collective ground state of the circuit, and $\hat{U}_{{\rm{enc}}}(\overrightarrow{\theta_i})$ is the unitary that encodes the normalized version of the field in the wavefunction at time $t_i$.

For the remaining terms of the PDE, we discretize the operators following the finite difference methods:
\begin{subequations}
\begin{eqnarray}
\frac{\partial}{\partial x}f(x,t) &\equiv& \frac{f(x_{k+1},t)-f(x_{k-1},t)}{2dx},\\
\frac{\partial^2}{\partial x^2}f(x,t)
&\equiv& \frac{f(x_{k+1},t)-2f(x_k,t)+f(x_{k-1},t)}{(dx)^2},~~\\
\frac{1}{2}\frac{\partial}{\partial x}f^2(x,t) &\equiv& f(x_k,t)\bigg[\frac{f(x_{k+1},t)-f(x_{k-1},t)}{2dx}\bigg],\\
\frac{\partial}{\partial t}f(x,t) &\equiv& \frac{f(x_k, t+dt)-f(x_k,t)}{dt},
\end{eqnarray}
\end{subequations}
notice that we have used the identity $\partial_{x}[f^{2}(x,t)]=2 f(x,t)\partial_{x}[f(x,t)]$. 

Assuming periodic boundary conditions, the implementation of spatial derivatives can thus be   decomposed in terms of the shifter $\mathbf{\hat{\mathcal{A}}}$ that shifts the field amplitude forwards or backwards. If we have the state $\ket{\psi^i}=\sum_{\ell}\psi^i_{\ell}\ket{\ell}$ the shifter displaces it following
\begin{equation}
     \mathbf{\hat{\mathcal{A}}}|\psi\rangle= \psi^i_{2^N}|0\rangle+\sum_{\ell=0}^{2^N-1}\psi^i_{\ell}|\ell+1\rangle,
\end{equation}
Then, the derivatives w.r.t. the position can be written as 
\begin{eqnarray}\label{eq:dx_operator}
\frac{\partial}{\partial x}=\frac{\mathbf{\hat{\mathcal{A}}}-\mathbf{\hat{\mathcal{A}}}^{\dag}}{2dx},\quad \frac{\partial^{2}}{\partial x^{2}}=\frac{\mathbf{\hat{\mathcal{A}}}-2\mathbf{\hat{\mathcal{I}}}+\mathbf{\hat{\mathcal{A}}}^{\dag}}{(dx)^2}
\end{eqnarray}

This scheme naturally implements periodic boundary conditions without any additional gate requirememts. Non- periodic boundary conditions can also be implemented by following the approach in \cite{over2025boundary}.

The non linear terms can be written in operator form as:

\begin{eqnarray}
\frac{1}{2}\frac{\partial}{\partial x}f^2(x,t) =\mathbf{\hat{\mathcal{D}}}\bigg[\frac{\mathbf{\hat{\mathcal{A}}}-\mathbf{\hat{\mathcal{A}}}^{\dag}}{2dx}\bigg],
\end{eqnarray}
where the nonlinear operator $\mathbf{\hat{\mathcal{D}}}$ is defined in such a way that it's action post measurement corresponds to a sum of products of the $\ell$th component of the shifted field with the $\ell$th component of the field.

The circuit implementation of $\mathbf{\hat{\mathcal{D}}}$ consists on the application of another ansatz $\hat{U}(\overrightarrow{\theta}_i)$  on $n$ auxilliary qubits followed by set of cascades of CNOTs gates between the algorithmic (control) and the auxiliary (target) qubits. These CNOTS will create copies of the basis state of the algorithmic qubits, $\ket{\psi}=\sum_{\ell}\ket{\ell}$ upon acting on $\bra{\mathbf{\bar{0}}}$. This way, upon measurement, we will have on the auxilliary qubits:

\begin{eqnarray*}&&
\bra{\mathbf{\bar{0}}}{\rm{CNOT}}_{\ell,\ell^{'}}\sum_{\ell^{'}}\psi_{\ell^{'}}\ket{\ell'}=\sum_{\ell,\ell'}\psi_{\ell} \braket{\ell}{\ell'},
\end{eqnarray*}

We have a non-zero entry when $\braket{\ell'}{\ell}=1$, that only occurs when $\ell'=\ell$ . Thus, by choosing the circuit implementing $\partial_{x}$ we apply this protocol to generate the non-linear term via the extra ancilla qubits, allowing for terms of the form $\sum_l \psi_{\ell} \partial_x \psi_{\ell}$ (in fact any power $\psi^{m}$ can be generated by using $m\times n$ ancilla qubits and taking the same strategy $m$ times.). 

The time derivative, obtained via the updating rule $f(x,t+dt) =  [ \mathbf{\hat{\mathcal{I}}}+ dt\mathbf{\hat{O}}(f(x,t))] f(x,t)$,  can be implemented in a variational algorithm through a cost function that compares the field at time $t+dt$ obtained through the encoding with the field obtained via the evolution 
\begin{equation}
\mathcal{C}(\lambda_{i+1},\overrightarrow{\theta}_{i+1})=||f(x, t_{i+1})-(\mathbf{\hat{\mathcal{I}}}+dt\mathbf{\hat{O}}) f(x, t_{i})||,
\label{cost_function}
\end{equation}
Notice that the cost function explicitly depends on the encoding angles at the nest time step $\overrightarrow{\theta}_{i}$ and the normalization factor $\lambda_i$ from the ansatz decomposition given in Eq.~(\ref{Encoding}). Then, the differential equation can be solved with an hybrid classical-quantum algorithm finding the parameters $\{\lambda_i,\overrightarrow{\theta}_{i}\}$ that minimizes the distance on the cost function in Eq.~(\ref{cost_function}).

\subsubsection*{Invisicid Burgers Equation}

For the inviscid Burgers equation we have $\nu=0$. This simplifies the cost function to:

\begin{eqnarray}
    \label{eq:cost_inv}
    \mathcal{C}(\lambda_{i+1},\overrightarrow{\theta}_{i+1})&=& |\lambda_{i+1}|^2+|\lambda_{i}|^2\\\nonumber
    &-&2\lambda_{i+1}\lambda_{i}[\mathcal{C}_1(\overrightarrow{\theta}_{i+1})+\mathcal{C}_2(\overrightarrow{\theta}_{i+1})],  
\end{eqnarray}

where

\begin{widetext}
\begin{subequations}
\begin{eqnarray}
\label{eq:Inv_Exp1}
\mathcal{C}_1(\overrightarrow{\theta}_{i+1}) &=&  \langle \mathbf{\bar{0}}| \hat{U}_{{\rm{enc}}}^{\dag}(\overrightarrow{\theta}_{i+1})\hat{U}_{{\rm{enc}}}(\overrightarrow{\theta}_{i})|\mathbf{\bar{0}}\rangle,\\
\label{eq:Inv_Exp2}
\mathcal{C}_2(\overrightarrow{\theta}_{i+1}) &=& \lambda_{i}\frac{dt}{2dx}\bigg[\langle \mathbf{\bar{0}}| \hat{U}_{{\rm{enc}}}^{\dag}(\overrightarrow{\theta}_{i+1})\mathbf{\Tilde{D}}\hat{U}_{{\rm{enc}}}(\overrightarrow{\theta}_{i})|\mathbf{\bar{0}}\rangle -\langle \mathbf{\bar{0}}| \hat{U}_{{\rm{enc}}}^{\dag}(\overrightarrow{\theta}_{i+1})\mathbf{\Tilde{D}}\mathbf{\hat{A}}^{\dag}\hat{U}_{{\rm{enc}}}(\overrightarrow{\theta}_{i})|\mathbf{\bar{0}}\rangle\bigg].  
\end{eqnarray}
\end{subequations}
\end{widetext}

where the norm can now be extracted as:
\begin{equation}
    \lambda_{i+1}= \lambda_{i}|[\mathcal{C}_1(\overrightarrow{\theta}_{i+1})+\mathcal{C}_2(\overrightarrow{\theta}_{i+1})]|
\end{equation}

The invicid equation has a closed solution $f(x,t)=(ax+b)/(at+1)$ for an initial condition $f(x,0)=ax+b$. Therefore, along the manuscript we have chosen as initial condition $f(x,0)=3x$ which will have a field solution $f(x,t)= 3x/(t+1)$.  Thus, it is possible to compare directly the result from the quantum processor with the analytical solution. This problem can still be solved using periodic boundary conditions by using a backwards finite difference method for the spatial derivative, for the only ill defined point would be at $x=0$, but $f(t,0)\frac{f(t,0)-f(t,0-dx)}{dx}=0$, given that $f(t,0)=0$.

Given that, for this example, the functional form does not change (it stays as a linear function for all time), after solving for the parameters $\overrightarrow{\theta}_{i+1}$, one can extract the norm and propagate it through multiple time steps. Figures \ref{fig:timevol} f and g demonstrate this evolution for 50 time steps.

\subsubsection*{Viscid Burgers Equation}

For the visicid Burgers equation, we consider as initial condition a sine wave, $f(x, 0)= \sin{(2\pi x)}$ and periodic boundary conditions. We focus on the turbulent regime with a viscosity of $\nu=0.01$.

 The cost function given in Eq.~(\ref{cost_function}) can be explicitly written as follows:

\begin{eqnarray}
    \mathcal{C}(\lambda_{i+1},\overrightarrow{\theta}_{i+1})&=& |\lambda_{i+1}|^2+|\lambda_{i}|^2\\\nonumber
    &-&2\lambda_{i+1}\lambda_{i}[\mathcal{C}_1(\overrightarrow{\theta}_{i+1})+\mathcal{C}_2(\overrightarrow{\theta}_{i+1})+\mathcal{C}_3(\overrightarrow{\theta}_{i+1})],  
\end{eqnarray}
where the contributions of the cost function $\mathcal{C}_k(\overrightarrow{\theta}_{i+1})$ are given by:
\begin{widetext}
\begin{subequations}
\begin{eqnarray}
\label{eq:Viscid_Exp1}
\mathcal{C}_1(\overrightarrow{\theta}_{i+1}) &=& \bigg[1-2 \nu \frac{dt}{(dx)^2}\bigg] \langle \mathbf{\bar{0}}| \hat{U}_{{\rm{enc}}}^{\dag}(\overrightarrow{\theta}_{i+1})\hat{U}_{{\rm{enc}}}(\overrightarrow{\theta}_{i})|\mathbf{\bar{0}}\rangle,\\
\label{eq:Viscid_Exp2}
\mathcal{C}_2(\overrightarrow{\theta}_{i+1}) &=& \nu \frac{dt}{(dx)^2}\bigg[\langle \mathbf{\bar{0}}| \hat{U}_{{\rm{enc}}}^{\dag}(\overrightarrow{\theta}_{i+1})\mathbf{\hat{A}}\hat{U}_{{\rm{enc}}}^{\dag}(\overrightarrow{\theta}_{i})|\mathbf{\bar{0}}\rangle+\langle \mathbf{\bar{0}}| \hat{U}_{{\rm{enc}}}^{\dag}(\overrightarrow{\theta}_{i+1})\mathbf{\hat{A}}^{\dag}\hat{U}_{{\rm{enc}}}(\overrightarrow{\theta}_{i})|\mathbf{\bar{0}}\rangle\bigg], \\
\label{eq:Viscid_Exp3}
\mathcal{C}_3(\overrightarrow{\theta}_{i+1}) &=& \lambda_{i}\frac{dt}{2dx}\bigg[\langle \mathbf{\bar{0}}| \hat{U}_{{\rm{enc}}}^{\dag}(\overrightarrow{\theta}_{i+1})\mathbf{\Tilde{D}}\mathbf{A}\hat{U}_{{\rm{enc}}}^{\dag}(\overrightarrow{\theta}_{i})|\mathbf{\bar{0}}\rangle -\langle \mathbf{\bar{0}}| \hat{U}_{{\rm{enc}}}^{\dag}(\overrightarrow{\theta}_{i+1})\mathbf{\Tilde{D}}\mathbf{\hat{A}}^{\dag}\hat{U}_{{\rm{enc}}}^{\dag}(\overrightarrow{\theta}_{i})|\mathbf{\bar{0}}\rangle\bigg].  
\end{eqnarray}
\end{subequations}
\end{widetext}

and the norm can now be extracted as:
\begin{equation}
    \lambda_{i+1}= \lambda_{i}|[\mathcal{C}_1(\overrightarrow{\theta}_{i+1})+\mathcal{C}_2(\overrightarrow{\theta}_{i+1})+\mathcal{C}_3(\overrightarrow{\theta}_{i+1})]|
\end{equation}

\subsection{Heuristic circuit optimization}\setcounter{subsection}{2}

In general, to generate any field $f(x,t)$, the encoder has to have entangling gates; depending on the complexity of the field and the connectivity of the hardware, the unitary $\hat{U}_{{\rm{enc}}}(\overrightarrow{\theta}_i)$ will be generated with all-to-all gates or not. As in current SC hardware, all-to-all connectivity has not been implemented yet, we use the \textit{hardware efficient anzatz}, where we only consider entangling gates between neighboring qubits to reduce the overhead of the large amount of swap operation needed to implement a long distance interaction.

The selection of $\hat{U}(\overrightarrow{\theta}_{i})$ has a deep impact in the convergence of the optimization. The rule of thumb for choosing an adequate anzatz is combining shallow circuits that are less prone to be affected by the noise, while being expressive enough to encode  the velocity field but with an optimal number of variational parameters to avoid \textit{barren plateaus}~\cite{mcclean2018barren}.

\subsubsection*{ Circuit simplifications and heuristics}
We further optimize gate depth across all operators by taking account of the following simplification available for Hadamard tests: we can classify every gate into 2 categories, gates whose action on the state $|0\rangle$ change this state and gates that do not. For the gates that do not change  $|0\rangle$, there is no need for the extra control on the Hadamard qubit.

\begin{equation}
    \label{eq:Hcont-simplifi}
    \begin{split}
        U_C(|0\rangle +|1\rangle)|0\rangle ^{\otimes n} &= |0\rangle |0\rangle ^{\otimes n} +|1\rangle U |0\rangle ^{\otimes n} =
        \\ =|0\rangle U|0\rangle ^{\otimes n} +|1\rangle U |0\rangle ^{\otimes n}&= (|0\rangle+|1\rangle)U|0\rangle^{\otimes n}
    \end{split}
\end{equation}
given that the Toffoli gate is such a gate, every Toffoli with a control on the Hadamard qubit can be reduced to a simple CNOT. This simplification significantly reduces the circuit depth. 

Additionally, we introduce modified controlled-Y rotations by removing one of the CNOTs used to establish such a rotation. Although this gate is not equivalent to a controlled-Y rotation, it still establishes a parameter space big enough to encode the solution and allows us to save an extra 2 qubit gate per rotation used.

To efficiently implement the adder we start from  Ref~\cite{lubasch2020variational}, where if we consider $N$ qubits to encode the field, the adder will require $(N-2)$ auxiliary qubits that need to be connected through $(N-2)$ CNOTS and $2(N-1)$ Toffoli gates. However this is not an optimal design. Starting from the previous simplification, all Toffolis connected to the Hadamard qubit can be simplified to a simple CNOT. Next, all Toffolis that are applied in pairs can be reduced to Phase-Relative Toffolis, which only required half of the number of 2-qubit gates to be implemented \cite{maslov2016advantages, motzoi2017linear} when compared to normal Toffolis. Therefore, if we use N qubits to encode the field, in our design, we will require $(N-2)$ auxiliary qubits, $(N+1)$ CNOTs, $2(N-3)$ relative phase Toffolis and only $\boldsymbol{1}$ Toffoli.

\subsubsection*{ Ansatz building block}

As previously mentioned, the function f is dicretized and encoded into a quantum circuit using a unitary $\hat{U}_{encod}$. This unitary represents our ansatz and it's construction is fundamental for VQA in the NISQ era. 
The underlying rules for ansatz construction given the heavy Hex Topology considered are then to limit the number of 2 qubit gates to a minimum and to establish 2 qubit gates between nearest neighbours, therefore avoiding long range CNOTs and the adding of costly SWAPS.

This way, our basic building block will be $Y(\theta)+CNOT+Y^{\dagger}(\theta)$.  We choose rotations over the y-axis because we demand real amplitudes for encoding the field $f(x,t)$.

\begin{figure}[!t]
    \centering
    \includegraphics[width=1\linewidth]{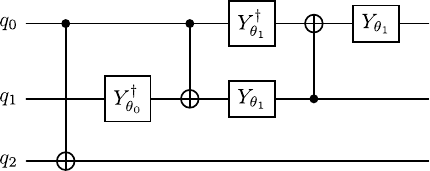}
    \caption{\textbf{Ansatz Used to Solve the Inviscid Burgers Equation}}
    \label{fig:InViscid_Ans}
\end{figure}

\begin{figure}[!b]
    \centering
    \includegraphics[width=1\linewidth]{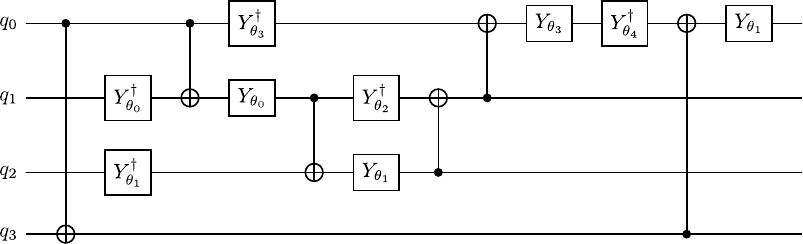}
    \caption{\textbf{Ansatz Used to Solve the Viscid Burgers Equation}}
    \label{fig:Viscid_Ans}
\end{figure}

We limit the number of circuit parameters to reduce the depth of the circuit as much as possible. This way, for the Inviscid Equation problem, we consider only 2 parameters and for the Viscid Equation problem we considered 5. Figures \ref{fig:InViscid_Ans} and \ref{fig:Viscid_Ans} show the ansatze considered, respectively.

Further research on how to efficiently build variational ansatze appropriate for quantum CFD problems is ongoing.

\subsection{Quality Metrics}\setcounter{subsection}{3}

Over the course of this paper we use 2 quality metrics to quantify how good our solutions are when compared to a classical evolution.

Naively, the first metric that one could use is the fidelity metric, defined as :

\begin{equation}
    F= |\langle \psi_1| \psi_2 \rangle |^2
\end{equation}

The fidelity will compare how close 2 quantum states, $|\psi_1\rangle$ and $|\psi_2\rangle$, are to each other, with $F=1$ when $|\psi_1\rangle=|\psi_2\rangle$.  This is a standard metric that allows us to quantify how good the parameters $\overrightarrow{\theta}_i$ that encode the discretized function into the unitary are.

However the fidelity is not enough given that it cannot quantify how accurate the normalizing parameter $\lambda_i$ is. Given this we also use the mean squared error:

\begin{equation}
    {\rm{MSE}} = \frac{1}{N}\sum_j (\lambda^{ideal}|\psi_j\rangle^{ideal} - \lambda|\psi_j\rangle)^2
    \label{eq:MSE}
\end{equation}

The MSE now quantifies the distance between the ideal solution and the obtained solution and allows us to compare norms.
We reconstruct the fields using a qiskit's state-vector simulator.

\subsection{Experiment}\setcounter{subsection}{4}

In this work we used IBM's quantum computers \cite{IBM_Machine}, readily available through the cloud. Specifically, the main results have used IBM's Marrakesh processor.

This machine uses fixed frequency qubits and tunable couplers to implement 2 qubit gates, ordered in a heavy hex lattice grid system. Although this degree of connectivity is subpar to the algorithm considered, this is compensated by the high 2 qubit gate fidelities achieved (above $99$\%) and high qubit lifetimes ($T_1$ and $T_2$ $\approx$ 100 $\mu$s). In addition, these superconducting systems allows for the execution of a high number of shots, an essential requirement for the running of variational algorithms, due to the optimization of the cost functions, and the application of error mitigation - every operator evaluation on the quantum computer was performed using 10000 shots. 

Both systems use as native gate set the gates $X$, $SX$, $RZ$, $RX$, $RZZ$ and $CZ$ requiring a high degree of transpilation from the circuit level implementation to the physical level. This was automatically provided by using qiskit transpilation pass managers \cite{javadi2024quantum}, at optimization level 3 to reduce the number of 2 qubit gates implemented to it's possible minimum.

Further details on the hardware specifications can be found in \cite{IBM_Machine}. 

\subsection{Error Mitigation and Supression }\setcounter{subsection}{6}

In order to successfully run complex experiments on existing hardware, error suppression and mitigation techniques are a critical necessity. 

To that end, the circuits in this article were implemented using Dynamical Decoupling \cite{PhysRevApplied.20.064027}, a powerful error suppression technique that aims to preserve the coherent state of a qubit by employing pulse sequences such that the average unwanted qubit interaction with the environment is zero. This technique is most beneficial when qubits undergo long periods of idling time -- exactly what happens with the Hadamard test implementation in this paper where, for example, the measured qubit starts the ansatz circuit but is not used to implement every controlled operation in the ansatz therefore remaining inactive.

Indeed throughout the examples here solved, we found the application of dynamical decoupling essential to obtain accurate results. Specifically, we applied dynamical decoupling with the pulse sequence ''XY4" \cite{PhysRevApplied.20.064027}, through the usage of qiskit's Sampler primitive \cite{javadi2024quantum}. This sequence has the advantage of being simple while universally suppressing all sources of local noise, experimentaly outperforming simpler decoupling sequences based on CPMG.

To further improve the accuracy of our results, we studied the usage of noise extrapolation protocols. Given the demanding shot usage and variance of the Hadamard test \cite{preti2025gradients}, we restricted our strategies to scalar rescaling techniques. Thus, we studied the effect of Digital Zero Noise Extrapolation (quadratic and exponential) and a new depolarizing noise estimation method specially tailored towards Hadamard circuits.

Zero Noise Extrapolation (ZNE) \cite{giurgica2020digital} is a well established technique that can be used to rescale an expectation value through error amplification. In ZNE the source of error is artificially increased and a function is fitted to the expectation values measured at different noise levels $\zeta$. In this work we take the standard approach of increasing noise digitally  by locally increasing the  number of 2 qubit gates -- the main source of error. Thus, for each circuit, we run 3 circuits with 1, 3 or 5 CNOTS, corresponding to $\zeta=1$, $\zeta=3$ and $\zeta=5$ respectively. After measurement, a function is fit and extrapolated to the zero noise level. In our work, we considered a quadratic regression and an exponential one, which allows us to compare different fitting strategies.

It has been demonstrated that ZNE can perfectly mitigate depolarizing noise \cite{giurgica2020digital}. However in real world settings, the increase of noise by multiple factors can often throw our expectation values into completely chaotic regimes, where the obtained response no longer works under the depolarizing noise channel. To avoid this problem, we take inspiration from more recent error mitigation protocols \cite{PhysRevLett.127.270502,PhysRevE.104.035309} where a depolarizing rate is estimated via specially designated circuits. In particular, code folding methods and longer circuits used to estimate the role of noise very often require circuits depths to increase into the information mixing regime, resulting in an inability to correctly invert the noise channel. 

Thus, we now introduce a new, streamlined technique based on a global depolarizing channel in the context of Hadamard test algorithms. The evolution of the density matrix under a global depolarizing noise can be described by:

\begin{equation}
    \rho \rightarrow (1-\eta)\rho_{ideal} +\frac{\eta}{2^n}\mathcal{I}
    \label{eq:noise_channel}
\end{equation}
where $\rho_{ideal}$ represents the ideal evolution without any noise. If one now applies a final Hadamard gate on the measured qubit, (considered ideally as a simple single qubit gate), the circuit will uncompute back towards a state where if the measured qubit is in $|0\rangle$ every qubit remaining qubit should also be at $|0\rangle$.

The application of this Hadamard gate will not change the expression from Eq.~(\ref{eq:noise_channel}). Therefore, if we measure all qubits, the ratio between counting all qubits at $|0\rangle$, namely $P_N(0)$, and the Hadamard one termed as $P_H(0)$, it is possible to estimate the depolarizing noise through the ratio 
\begin{equation}
\frac{P_N(0)}{P_H(0)}
=
(1-\eta)+\frac{\eta}{2^n},
\label{eq:lambda_frac}
\end{equation}

where $\eta$ is the depolarizing probability and $n$ is the number of qubits. Given that the quantity of interest in a Hadamard test is obtained via the difference between the counts of the hadamard qubit at $|0\rangle$ and $|1\rangle$ and the identity $\mathcal{I}$ will affect both results in the same way, the mitigated expectation value is obtained by rescaling the noisy value by $1/(1-\eta)$. That is:
\begin{eqnarray*}
\langle O \rangle_{noisy} &=& \Tr(|0\rangle\langle 0| \rho)- \Tr(\ketbra{1}{1} \rho)\\ 
&=& (1-\eta) [\Tr(|0\rangle\langle 0| \rho_{ideal})- \Tr(\ketbra{1}{1} \rho_{ideal})]\\ &+&\frac{\eta}{2^n}[\Tr(|0\rangle\langle 0| I)- \Tr(\ketbra{1}{1} I)]\\
&&=(1-\eta)\langle O \rangle_{ideal}
\end{eqnarray*}
and therefore:

\begin{equation}
    \langle O \rangle_{ideal} = \frac{\langle O \rangle_{noisy}}{(1-\eta)}
\end{equation}

Using the calculation of the norm for a single time step as our guiding principle, we compare quadratic ZNE, exponential ZNE and Hadamard Verification. As can be seen from Fig.~\ref{fig:Time_March_Mit}b, depolarization inversion outperforms the other methods considered and was therefore used. This method has a 2-fold advantage: first it requires less circuits to be run (from 3 to 2) and second all circuits run are of the same depth of the circuit of interest. Therefore, because it targets the same noise channel that zero noise extrapolation mitigates without incurring extra depth, it achieves better results. This however is predicated on the validity of the global depolarizing noise channel. In case this channel is not valid or does not approximate the noise from the real device, the method will break down.

At the same time, one can test the quality of the circuits being run on the machine by looking at $(1-\eta)$. A high value of $(1-\eta)$ will be associated with circuits of shallower depth and better results, a lower value to circuits of deeper depths with more error. Further research is required to incorporate other more advanced noise models .

\bibliography{References}

\end{document}